# Ultra-sensitive Flexible Sponge-Sensor Array for Muscle Activities Detection and Human Limb Motion Recognition

Jiao SUO[1,#], Yifan LIU[1,#], Clio CHENG[2], Keer WANG[1], Meng CHEN[1,*], Ho-yin CHAN[1,*], Roy VELLAISAMY[3], Ning XI[4], Vivian W. Q. LOU[2], and Wen Jung LI[1,*]

[1]Dept. of Mechanical Engineering, City Univ. of Hong Kong, Hong Kong, China
[2]Dept. of Social Work and Social Adm., The University of Hong Kong, Hong Kong, China
[3]James Watt School of Engineering, University of Glasgow, Scotland, United Kindom
[4]Dept. of Industrial and Manufacturing System Engineering, The University of Hong Kong, Hong Kong, China

**Abstract**

Human limb motion tracking and recognition plays an important role in medical rehabilitation training, lower limb assistance, prosthetics design for amputees, feedback control for assistive robots, etc. Lightweight wearable sensors, including inertial sensors, surface electromyography sensors, and flexible strain/pressure, are promising to become the next-generation human motion capture devices. Herein, we present a wireless wearable device consisting of a sixteen-channel flexible sponge-based pressure sensor array to recognize various human lower limb motions by detecting contours on the human skin caused by calf gastrocnemius muscle actions. Each sensing element is a round porous structure of thin carbon nanotube/polydimethylsiloxane nanocomposites with a diameter of 4 mm and thickness of about 400 μm. Ten human subjects were recruited to perform ten different lower limb motions while wearing the developed device. The motion classification result with the support vector machine method shows a macro-recall of about 97.3% for all ten motions tested. This work demonstrates a portable wearable muscle activity detection device with a lower limb motion recognition application, which can be potentially used in assistive robot control, healthcare, sports monitoring, etc.

# Introduction

Human lower limb motion tracking and recognition have been important research topics because of their applicability in medical rehabilitation training [1], lower limb motion assistance [2], gait analysis (which are useful for disease diagnosis) [3], prosthetics design for amputees [4], entertainment virtual reality (VR) games [5], assistive robotics, etc. For example, it is extremely critical in human-machine interaction technologies and could allow machines to serve humans better by providing limb motion information as feedback information to actuate assistive robotic systems more safely and efficiently. Optical systems based on cameras is one of the most used methods to track and recognize limb motions [6-10]. However, the vision-based technology requires proper conditions of the environment for effective implementation and stable results, i.e., surrounding lighting condition, the locations of the cameras, occlusion limit, etc. [8]. In addition, the data size of the vision-based recording is large that sufficient computational power is required in processing the collected data. These disadvantages limit the vision-based technology in gaining more applications in daily life. On the other hand, angle encoders are often used to measure the bending angle of hip joint or knee joint, and hence, it is also a commonly used method to capture human limps motions [7, 11]. However, it is not very compliant with the human body, i.e., is prone to cause discomfort and even joint injury during long-term usage. More recently, small and lightweight wearable sensors have been developed to perform human motion capturing and recognition. Devices based on inertial sensors (e.g., accelerometer, gyroscope, inertial measurement unit (IMU)) are the most used in both laboratory research [12-15] and commercial products (e.g., Apple Watch [16], Fitbit smartwatch [17], etc.). Inertial sensors measure acceleration and angular velocity of objects in motion with three mutually perpendicular axes, so they are placed on each joint to capture fine movements of the human limbs, including unwanted vibrations (e.g., see [18, 19]).

In addition to tracking limb motions by measuring motions of the human joints, muscles activities (i.e., stiffening, relaxation, contraction, etc.) are also integral to human motions which are achieved with a complex and highly coordinated mechanical interaction between bones, muscles, ligaments and joints within the musculoskeletal system [20]. Therefore, measuring muscle activities has become a popular way for human motion capturing and recognition. Electromyography (EMG) [21, 22] and

Surface Electromyography (sEMG) [18, 23-27] acquire the electrical signals directly from muscles and has been used to record human lower limb motions -- good recognition accuracy has been achieved for several common motions. The signal amplitude of EMG is related to  sEMG is widely used due to its advantages of non-invasiveness and easy operation. In addition, EMG/sEMG are also used by combining them with inertial sensors to obtain more information [28, 29]. For instance, Shi et.al., achieved the recognition of 94.83% of six lower limb motions (i.e., horizontal walking, crossing obstacles, standing up, downstairs, go upstairs, and stop-rest) using three pairs of sEMG placing on thigh semitendinosus, lateral thigh muscle, and calf gastrocnemius [23]. Wang et.al., used five pairs of EMG placing on tibialis anterior muscle, gastrocnemius muscle, biceps femoris muscle, lateral femoris muscle, and tensor fascia muscle to recognize the motions of normal walking, uphill walking, downhill walking, squat, side squat and the mean accuracy got about 93.35% [21]. Zhou et.al., combined the data from three pairs of sEMG and one accelerometer, which were placed on the tibialis anterior, gastrocnemius medialis, soleus, and ankle joint, and showed recognition accuracy of dorsiflexion, plantar flexion, eversion, and inversion of foot of about 94.39% [28]. Usually, more than one piece of muscle activities needs to be measured in order to track/recognize limb motions, i.e., multiple sensing electrodes and/or inertial sensors are required across a subject's body and limbs.  Moreover, the signal from the EMG/sEMG sensors are known to be influenced by muscle fatigue (i.e., producing more noise) and hence could produce inconsistent results over time. This phenomenon decreases the sensor reliability during application, and the recognition accuracy could decrease about 5% after muscle fatigue [22]. Therefore, existing methods of measuring muscles activities should be improved for limb motion tracking and recognition applications.

Flexible and wearable pressure/strain sensors used to measure different human body signals, such as arterial pulse [30-33], joint bending [33-38], eyeball movements [33, 39, 40], throat vibration [41, 42], etc., have been attracting enormous interests during the last decade. Moreover, the combination of Internet of Things (IoT) technology and flexible sensor/sensor arrays has created more promising applications recently [5, 43-47].  Nevertheless, current research efforts mostly focus on the detection of motions around small areas of skin, i.e., collecting spatial information for capturing complex motions involving large muscles such as human lower limb motions are currently non-existent. Recently, the pressure distribution on the soles of the feet caused by different motions was detected using triboelectric nanogenerators (TENG) and then recognized [5, 47]. It is an indirect way to measuring limb motions compared to measuring calf muscles activities directly. Calf muscle consists of two main muscles, i.e., the gastrocnemius, and the soleus. Gastrocnemius muscle is the most superficial two-joint muscle which connects knee and ankle on the human leg, while soleus is a flat, wide muscle which sits slight deeper than the gastrocnemius, and it only crosses the ankle joint [48]. So, the gastrocnemius is used extensively during various limb motions since it connects to two joints and it works to flex the ankle and propels human forward when walking, running, jumping, or climbing stairs [49]. The failure of the medial gastrocnemius muscle reduces the peak ankle dorsiflexion angle after landing by 35.5% [50]. Therefore, gastrocnemius muscle is an ideal muscle for detecting multi human lower limb motions and it is also widely used in the sEMG method to decipher muscle activities [21, 23, 28]. This work proposes a portable device using a thin sponge-based 16-channel pressure sensor array to capture and recognize multi human lower limb motions by attaching the device to the skin around the gastrocnemius muscle. Hence, the data of the different deformations of the muscle due to different limb motions could be recorded and then processed by machine learning algorithms to recognize 10 different leg motions. Different from the sEMG sensors, which measure electrical activities of muscles, the thin sponge-based sensor array measures the muscles movements directly. Only one muscle (gastrointestinal muscle) and one sensing device are needed using our proposed technique for recognizing lower limb motions. The sponge-based sensor array has a weight of about 0.5 g and the circuit board (5.4 cm x 3.4 cm x 2.5 cm) has a weight of 58 g, which is very convenient to wear for human subjects.

## Results
### Device preparation

The entire device consists of a piece of 16-channel sponge-based pressure sensor array with the sensing area of 32 mm x 28 mm and a ESP32-based wireless signal acquiring/transmission circuit board, as shown in **Fig. 1**. The sensor array (with 4x4 elements) is extremely flexible and conformable to human skin (i.e., with each sensing element suspended and decoupled from other elements) to detect the human muscle deformation. A coated Parylene film around the sponge-based sensing elements enables the sensors to resist possible sweat and has been shown to be biocompatible by many past researchers. Sensors based on the sponge structure have been shown to exhibit high sensitivity for skin movement/vibration detection [51]. The wireless MCU board based on ESP32 was used to sample and transmit data to a computer via WIFI protocol. The sampling rate of the entire 16 channels is about 60Hz and the sampling interval for each channel is about 0.001s; this sampling rate can cover all the human motions according to the Nyquist sampling theorem, since the frequency of human activities is between 0~20Hz, and 98% of them is below 10Hz [55].

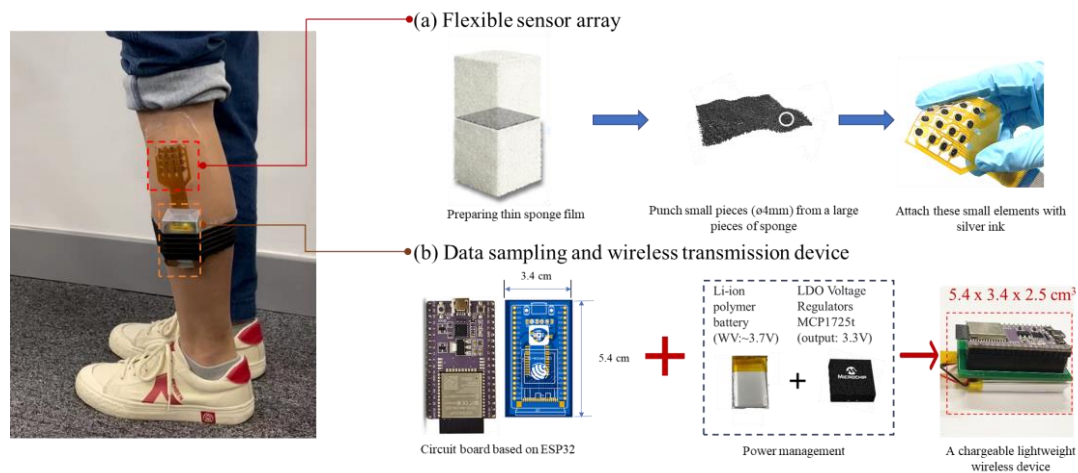

**Fig. 1. Structure of the whole portable motion detection device**. (a) Fabrication of the 16-channel sponge-based pressure sensor array. (b) Wireless circuit for data sampling and transmission.

**Muscle deformation detection principle**

The mechanism of the sensor array in detecting human muscle deformation is shown in **Fig. 2**(a); the movement of muscles induce the deformation of the sponge-based sensing elements which worked as the piezoresistive pressure sensors. A simple voltage dividing circuit (already integrated in the above wireless circuit board) was used to measure the sensor response, and voltage value was read as the output signal. Response of the individual sensory element is shown in **Fig. 2**(b); the rate of change of the output voltage increases with the applied pressure. The results show that the sensing element has a higher sensitivity (higher slope) in the lower pressure range of below 10kPa. **Fig. 2**(c) shows the signal output of all 16 channels of the sensor array when performing the motions of "lift heel" and "lay down". Each channel has different response to the motions as they measured the activities of different muscle positions. **Fig. 2**(d) presents the basic distribution of the response of the sensor array by calculating the mean value of the signal change rate of different channels. Therefore, the different pressure distribution caused by muscle activities due to different limb motions can be detected with the sponge sensor array.

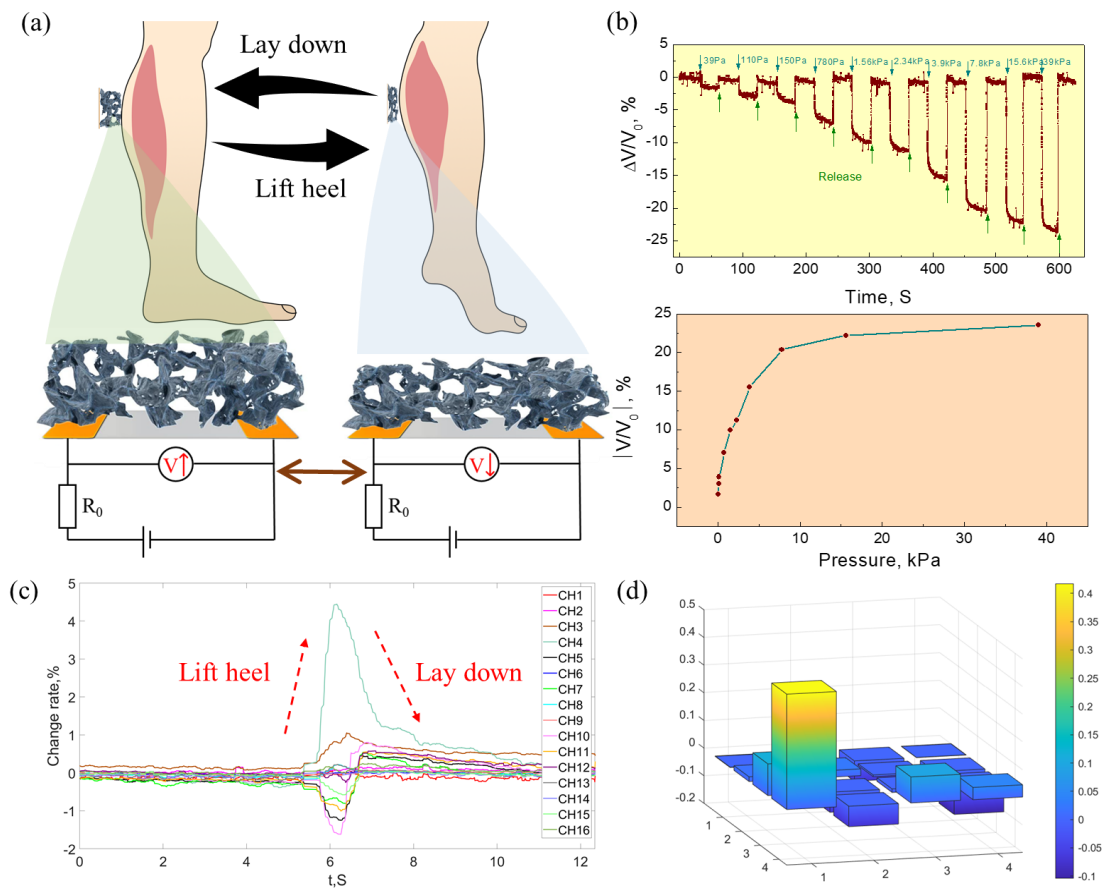

**Fig. 2. Response of the sensor array to pressures.** (a) Mechanism of sponge-based sensor detecting muscle deformation. (b) Response of individual sensory element to different applied pressure. (c) Response of all sixteen channels to the motions of lift hell and lay down with time. (d) Distribution of the mean value of rate of signal output change of the sensor array.

**Muscle activities detection with the sensor array**

The wearable sensor array-based device shown above has the advantages of excellent flexibility due to the hollow structure electrodes, soft sponge sensing materials, and the convenience of muscle deformation signal gathering using wireless data transmission. We have used the sensor array to detect 10 different lower limb motions, i.e., 1) sitting motions of lift heel, 2) lift toes, 3) foot inversion, 4) stretch leg forward, 5) backward and 6) standing motions of step in situ, 7) standing with foot inversion and 8) half-turn in situ (leg with sensory device as the axis), 9) walk forward and 10) walk backward. These lower limb motions are shown in the pictures of **Fig. 3**. These motions include *muscle training activities* (such as lift heel, lift toes, and stretch leg forward), *normal/daily walking activities* (such as step in situ, walk forward), also the *abnormal motions* (such as foot inversion) which are typical motions that may cause injury. The flexible sensor array was directly attached to the subjects' skin around the calf gastrocnemius muscle, and the signal processing box attached to the sensing array was fixed to the leg with an elastic strap to make sure it would not move or fall during specified actions of the subjects.

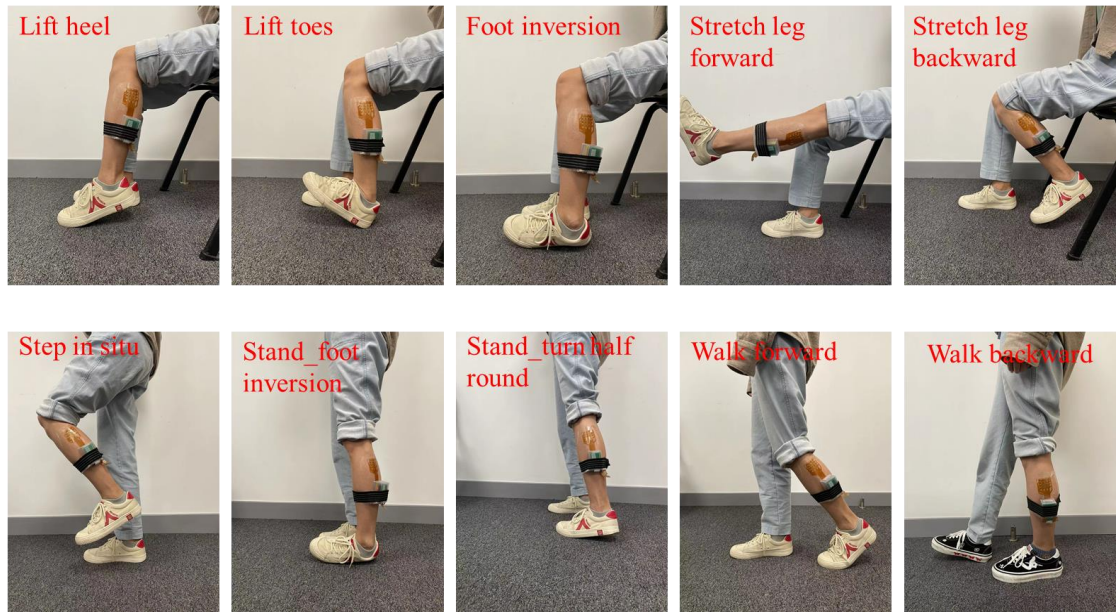

**Fig. 3. Pictures of the ten lower limb motions recognized by the flexible sensor array.**

Real-time response of all 16 channels to the ten different motions is shown in **Fig. 4**(a), which suggests each channel responds differently to the motions. Small movements such as foot inversion induced smaller response since smaller muscle motions were caused. On the other hand, big movements such as walking forward induced larger response because of the larger muscle motions. Separate signals collected from each of the channels are shown in **Fig. 4**(b). Also, the standard deviation values of the stationary state and motion state of multi-motions were calculated and compared in **Fig. 4**(c). Each channel changes more dramatically during motion activities than at rest, and the fifth channel shows the most obvious change during of the ten different motions.

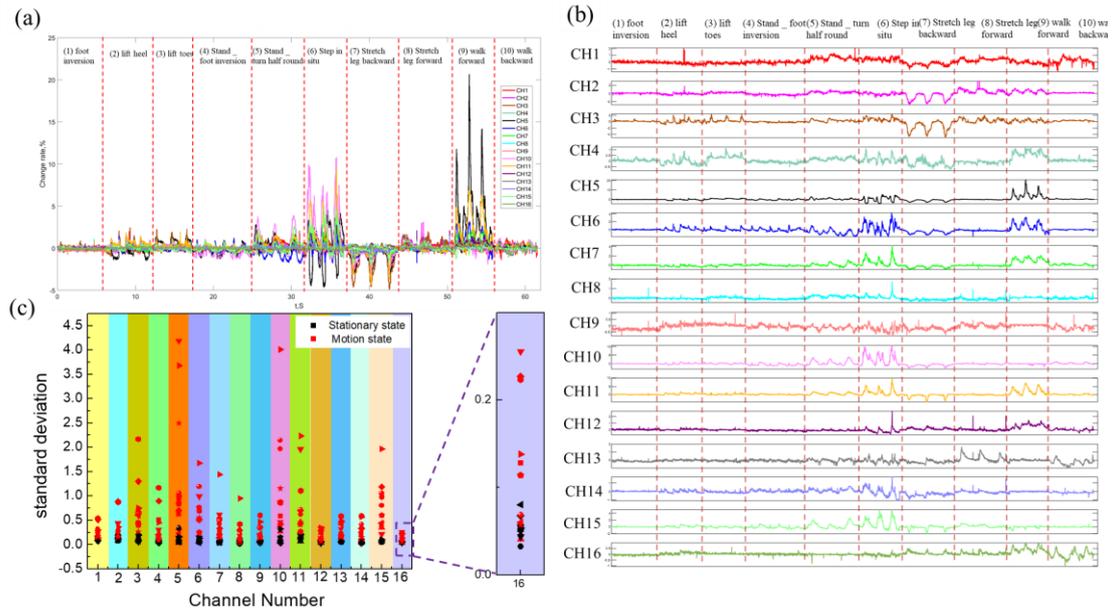

**Fig. 4. Response of the 16 channels of sensor elements to the ten lower limb motions**. (a) Real-time response of sensor array to multi-motions. (b) Response of each single channel to multi-motions. (c) Comparison of the standard deviation values of stationary state and motion state of the 10 different motions.

**Recognition of the lower limb motions**

Specific lower limb motion recognition has great value in health monitoring, rehabilitation, assistive robotic control, game entertainment, etc. Different from the widely used human limb motion tracking sensors such as camera systems, IMU sensors, sEMG, etc., our flexible pressure sensor array-based portable device provides rich motion information related to muscle activities and has been demonstrated to be a very reliable sensing device in recognizing many different lower limb motions. Support vector machine (SVM) is a classical classification method and works well with even small datasets, so it was selected to perform the recognition task based on the motion data from ten human subjects. The data process flow is shown in **Fig. 5**(a). Different segmentation methods, including fixed window and sliding window with different overlap were applied for pre-process. For each window, four time-series features (i.e., mean value, standard deviation, energy, and root mean square) of each channel were calculated as the input matrix to SVM. **Fig. 5**(b) show the classification results of ten different lower limb motions for ten human subjects, which has a macro-recall of 97.3%. The result indicates that high precise motion recognition can be achieved with motion data from the developed wearable device and simple classification method of SVM.

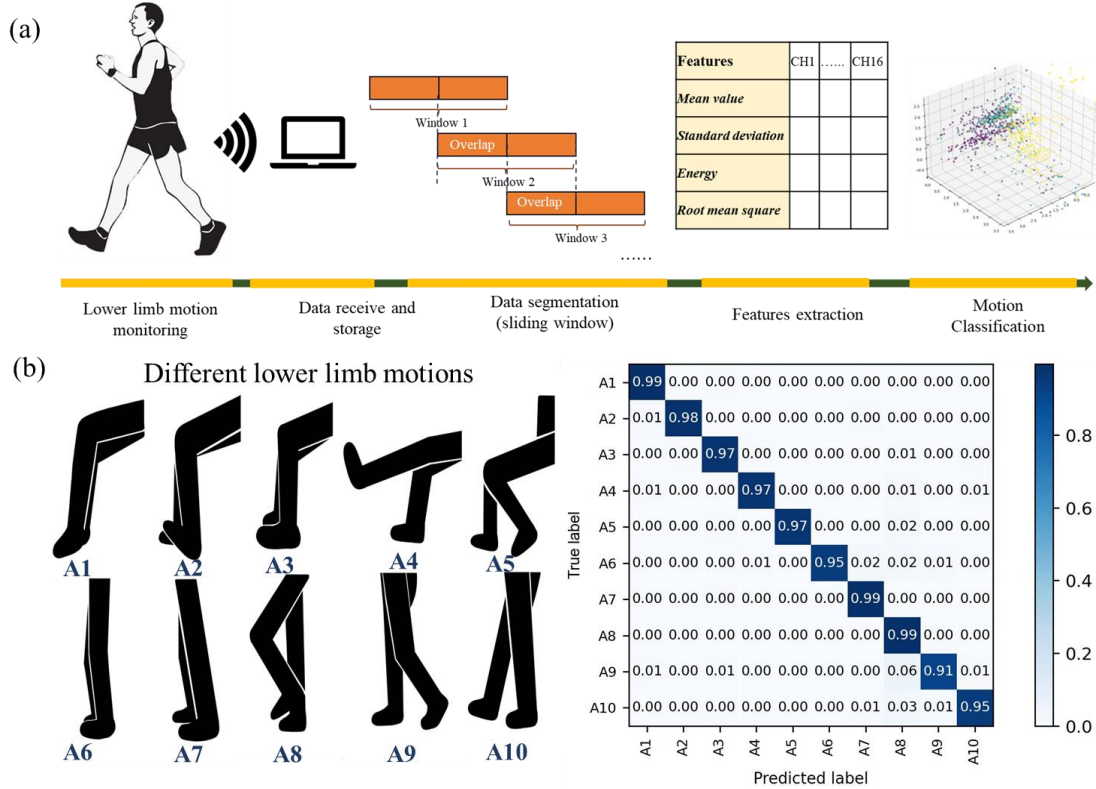

**Fig. 5. Recognition of ten kinds of different lower limb motions.** (a) Data process flow. (b) Classification results of ten different lower limb motions from ten human subjects (A1: lift heel, A2: lift toes, A3: foot inversion, A4: stretch leg forward A5: stretch leg backward, A6: standing with foot inversion, A7: turn round A8: step in situ, A9: walk forward, A10: walk backward).

**Health assessment monitoring**

The developed device can also be used to assess the leg health status by monitoring the standard assessment tests. Gait analysis is an essential part of diagnosis of neurologic disorders, assessment of the progress of rehabilitation, disease process of the lower limb, etc. [52]. Results of human gait analysis is shown in **Fig. 6** (a). The parameters of gait cycle, stance phase (each foot performs ground contact), swing (each foot performs lifting off the ground) and cadence are shown in **Fig. 6** (d), which are helpful for gait and related health analysis. The proportions of the stance phase and swing phase of a whole gait cycle detected by our developed device are similar to other researchers' results using other methods. For example, the stance phase lasts ~ 60% and swing phase lasts about ~ 40% of a whole gait cycle [53]. We have also run "thirty-second chair stand test" with several subjects. This is a test where a subject needs to standing up from a sitting position as many times as possible within 30 seconds, and is to assess the leg strength and endurance based on a score obtained by counting the number of stands within 30 seconds [54]. The output signals of the sit-to-stand test is presented in **Fig. 6**(b), and the number can be easily counted according to the results. On the other hand, the "tandem stance" is clinical way to assess the standing balance by considering postural steadiness in a heel-to-toe position with a temporal measurement [55]. A continuous recording of the tandem stance balance test by our device shows that it could record the time when the balance is completely lost due to the feet movement, and also the time when the body starts to shake before a subject loses balance (**Fig. 6**(c)). These signals provide rich information can could be potentially used by doctors or individuals to enhance health management methods.

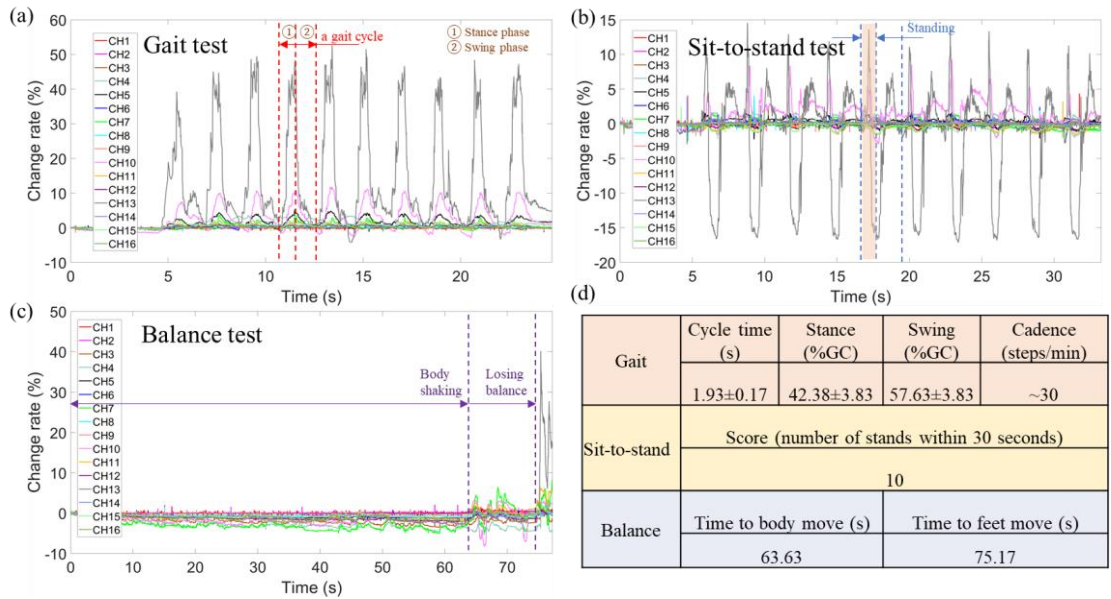

**Fig. 6. Results of health assessment tests.** (a) Gait analysis. (b) 30-second sit-to-stand test. (c) Tandem stance balance test. (d) Quantization parameters obtained according to the recorded test results.

## Discussion and conclusion

Various technology and sensing devices have been developed during the past two decade for human motion capture and recognition, especially with the emergence of smart wearables such as smart watch, smart wristband, etc. Currently, IMUs are the most used sensors in these wearable devices, whose working principle is to measure the angular and acceleration related parameters that are mainly dependent on the joint motions. On the other hand, muscle activity is an essential parameter understand in order to track and recognize complex human motions. Physiological signals from the muscles provide motion information from a totally different perspective. Currently surface electromyography (sEMG) sensors are the most used for muscle activity detection, but they face the problems of noises from muscle fatigue, poor portability, signal instability, etc. The wearable muscle activity detection device presented in this work has the advantage of good portability (simply wearing one sensor array on a designated skin area) and accurate motion recognition (over 90% accuracy can be achieved with ten different lower limb motions). This novel sensing device could potentially play a major role in various health-related fields by directly measuring skin deformation due to muscle activities.

In conclusion, a portable AI-enabled muscle activities detection device is developed for human limb motion recognition. The flexible 16-channel sensor array was uniquely designed to have suspended and mechanically decoupled sensing elements, each made with ultra-thin CNT/PDMS sponge nanocomposite. Ten different lower limb motions were recognized by wearing the developed device on the gastrocnemius of subjects (we have thus far tested ten human subjects). The SVM classification shows a macro-recall of 97.3% for ten motions of ten subjects. In addition, health-relate tests such as gait analysis can be also performed with the developed sensor array.

## Methods

**Sponge-based pressure sensor array preparation**

The sensing elements of the sensor array is made of the nanocomposite material carbon nanotube/polydimethylsiloxane (CNT/PDMS) which used isopropyl alcohol (IPA) as the solvent. Multiwall CNTs with diameters of 10–20 nm and lengths of 10–30 μm (provided by the manufacturer) were used. First, 0.5g MWCNTs were dispersed in a sufficient quantity of IPA and ultrasonicated for 20 min to obtain a dispersion of CNTs. Then, 10g PDMS-A base was added into the dispersion and ultrasonication for 10 min. Subsequently, the mixture was placed on a hotplate (IKA, Germany) maintained at 55°C to completely evaporate the IPA. Thereafter, PDMS-B agent (the weight ratio of PDMS-A and PDMS-B is 10:1) was added to the solution and mechanically mixed. Finally, air bubbles were removed from the mixture through vacuum treatment. The CNT/PDMS solution was then spread on the surface of a cube sugar to obtain the sponge structure. The sugar template was dissolved after the solution was cured in an oven at 70°C for about 2h to get the pieces of thin sponge film with the area of about 2cm x 2cm and thickness of about 400μm. The individual round sensory elements with the size of φ4mm x 400μm were then punched out. The flexible substrate with a thickness of ~ 150μm was fabricated with Polyimide (PI) as the structural material which covers a thin metallic conduction layer as electrical connectors. The individual sponge sensory elements were attached to each electrical connector of the flexible substrate with silver ink. Finally, a layer of 30nm-thick Parylene-c was coated on the surface of the entire sensor array.

**Signal acquisition and wireless transmission circuit board**

A simple voltage divider circuit was designed based on the feature rich MCU ESP32 board. A 16-channel multiplexer was used to manage the output voltage signal acquisition of the 16-channel sensor array. The acquired data were transmitted to a computer via WIFI protocol with the sampling rate of about 60Hz for each channel. The entire device was powered by a chargeable Li-ion polymer battery and a low-dropout regulator (LDO) chip MCP1725T.

**Experiments on human subjects**

Ten human subjects were recruited to perform 10 different lower limb motions including sitting state (i.e., foot inversion, lift heel, lift toes, stretch leg forward and stretch leg backward) and standing state (i.e., foot inversion, turn round, step in situ, walk forward and walk backward) with wearing the prepared portable device. The sponge-based sensor array was attached to human leg of gastrocnemius (head) with 3M Tegaderm transparent film, and the data acquisition and transmission board was fixed to human legs with elastic straps. The subjects were asked to perform each motion repeatedly with the frequency of about 0.5Hz and last for 90 seconds each time. Also, each subject had to perform 4 sets of each motion, with a minute of rest in between each set. The experiments were done indoors with stable WiFi coverage and data were received with a computer.

**Data process**

The relative rate of change (i.e., $X = (V-V_0)/V_0$) of the output voltage signal was first calculated. Then, segmentation with fixed window and sliding window were performed. The segmentation time varied in 2 s, 4 s, and 6 s, while the overlap for sliding window varied in 25%, 30%, 50% and 60%. The method which produces the best result (highest recognition accuracy) was applied. The average, root mean square, standard deviation, and the signal energy values of each channel for each window were calculated as the features as shown below:

$$Average: \bar{X} = \frac{1}{n}\sum_{i=1}^{N} X_i$$

$$\text{Root mean square}: RMS = \sqrt{\frac{1}{n}\sum_{i=1}^{N} X_i^2}$$

$$\text{Standard Deviation}: \sigma = \sqrt{\frac{1}{N}\sum_{i=1}^{N}(x_i - \bar{x})^2}$$

$$\text{Signal energy}: E = \frac{1}{n}\sum_{i=1}^{N}|X_i^2|$$

Principal component analysis (PCA) was used to reduce the dimension and compute the main components of all the features. Support vector machine (SVM) algorithm was applied to perform the classification (50% data for training while 50% for testing) and the metric recall was selected to evaluate its performance.

$$recall_i = \frac{N_{TP}(i)}{N_{TP}(i) + N_{FN}(i)}$$

A high recall value indicates that most of the behavior samples are correctly classified. Due to the size differences between the ten activities, we adapted the average recall value (macro-recall) as the main metric to represent the overall performance of the model, which is expressed as:

$$Macro\text{-}recall = \frac{1}{N}\sum_{i=1}^{N} Recall_i,$$

where $N$ is the total number of activities.

**Funding Source Acknowledgement**

This work was partially supported by funding from Hong Kong Research Grants Council: (1) Theme-based Research Scheme Project No. T42-717/20-R and (2) General Research Fund Project No. 11210819.

## Author Contributions

W. J. L., H.-Y. C. and J. S. conceived the initial concept for creating the sponge-based sensing array for muscle activities monitoring. J. S. developed the sensing array and performed the experiments. J. S. and Y. L. performed the data analysis; Y. L. specifically focused on the data using machine learning algorithms. J. S. and Y. L. contributed equally to this work. M. C., and Y. L. contributed to the circuit design; M. C. specifically developed the wireless data transmission solution. W. J. L. and H.-Y. C. supervised and guided the project. Y. L., K. W., and C. C. helped with the human subjects' test protocols and experiments. J. S., Y. L., and W. J. L. co-drafted this paper. V.A.L., N. X., and V. W. O. L. provided critical technical advice and comments for the experimental studies and helped revised the manuscript.